\documentclass[aip,showpacs,twocolumn,10pt]{revtex4}%
\usepackage{amsmath}
\usepackage{amsfonts}
\usepackage{amssymb}
\usepackage{graphics}
\usepackage{graphicx}%
\providecommand{\U}[1]{\protect\rule{.1in}{.1in}}
\draft
\begin{document}
\title{Directional amplification with a Josephson circuit}
\author{Baleegh Abdo}
\email{baleegh.abdo@yale.edu}
\author{Katrina Sliwa}
\author{Luigi Frunzio}
\author{Michel Devoret}
\affiliation{Department of Applied Physics, Yale University, New Haven, CT 06520, USA.}
\date{\today}

\begin{abstract}

Non-reciprocal devices, which have different transmission coefficients for propagating waves in opposite directions, are crucial components in many low noise quantum measurements. In most schemes, magneto-optical effects provide the necessary non-reciprocity. In contrast, the proof-of-principle device presented here, consists of two on-chip coupled Josephson parametric converters (JPCs), which achieves directionality by exploiting the non-reciprocal phase response of the JPC in the trans-gain mode. The non-reciprocity of the device is controlled in-situ by varying the amplitude and phase difference of two independent microwave pump tones feeding the system. At the desired working point and for a signal frequency of $8.453$ GHz, the device achieves a forward power gain of $15$ dB within a dynamical bandwidth of $9$ MHz, a reverse gain of -6 dB and suppression of the reflected signal by $8$ dB. We also find that the amplifier adds a noise equivalent to less than one and a half photons at the signal frequency (referred to the input). It can process up to $3$ photons at the signal frequency per inverse dynamical bandwidth. With a directional amplifier operating along the principles of this device, qubit and readout preamplifier could be integrated on the same chip.

\end{abstract}

\maketitle

\newpage

\section{Introduction}

Reciprocity is one of the basic symmetries in wave physics, in which the source and the detector can be exchanged without changing the transmission coefficient \cite{QED}. Breaking this symmetry (non-reciprocity) is particularly useful in amplification. 
Indeed, the signal source often needs to be protected from noise coming down the amplification chain in reverse. Hence, low noise measurement chains incorporate non-reciprocal devices such as circulators and isolators. These devices exploit a particular non-reciprocal effect known as Faraday rotation, which relies on ferrites and permanent magnets, in order to distinguish between polarized waves propagating in opposite directions \cite{Pozar}.
 
Circulators and isolators also play a pivotal role in state of the art quantum readout of superconducting qubits that utilize Josephson parametric amplifiers. Incorporating these devices as preamplifiers preceding the standard high electron mobility transistor amplifier (HEMT) has significantly improved signal-to-noise ratio. It has allowed performing single-shot quantum nondemolition readout of the qubit state, monitoring its quantum trajectories in real time and studying back-action effects on the qubit state due to weak and strong measurements \cite{QuantumJumps,RabiVijay,QubitJPC,korotkov}. However, present Josephson parametric amplifiers operating at the quantum limit, such as the Josephson bifurcation amplifier (JBA) \cite{JBA,VijayJBAreview}, the Josephson parametric amplifier (JPA) \cite{CastellanosAPL,Yamamoto} and the Josephson parametric converter (JPC) \cite{JPCnature,Jamp,JPCreview}, suffer from a serious limitation. They amplify in reflection even when the device has two ports. Thus, in order to separate between incoming and outgoing signals traveling on the same transmission line, and to protect the qubit from the amplified reflected signal, it is imperative to add a chain of at least two circulators between the qubit and the preamplifier. In the case of the JBA for example, which lacks a spatial and temporal separation between the pump and the signal, these circulators are further needed in order to reject the reflected strong pump. Unfortunately, using such a readout scheme comes at the cost of introducing losses between the cavity-qubit system and the preamplifier, hence adding noise to the processed signal. It also prevents an on-chip integration of the cavity-qubit system with the preamplifier.

For these reasons, there is a growing need for a directional Josephson
amplifier which (1) has two ports and works in transmission, (2) is matched to the input,
(3) treats the input and output fields differently (non-reciprocal), and (4) builds on the present performances of
Josephson parametric amplifiers, i.e. power gain, dynamical bandwidth, added noise and maximum input power \cite{JPCreview}.

While the transmission requirement might have a possible engineering solution,
for instance, by measuring the idler port in the case of the JPC (which is at
a different frequency), or using the two-port version of the JPA \cite{CastellanosNat}, matching to the input or achieving
non-reciprocity is a pressing challenge. 

\begin{figure}
[htb]
\begin{center}
\includegraphics[
width=\columnwidth 
]
{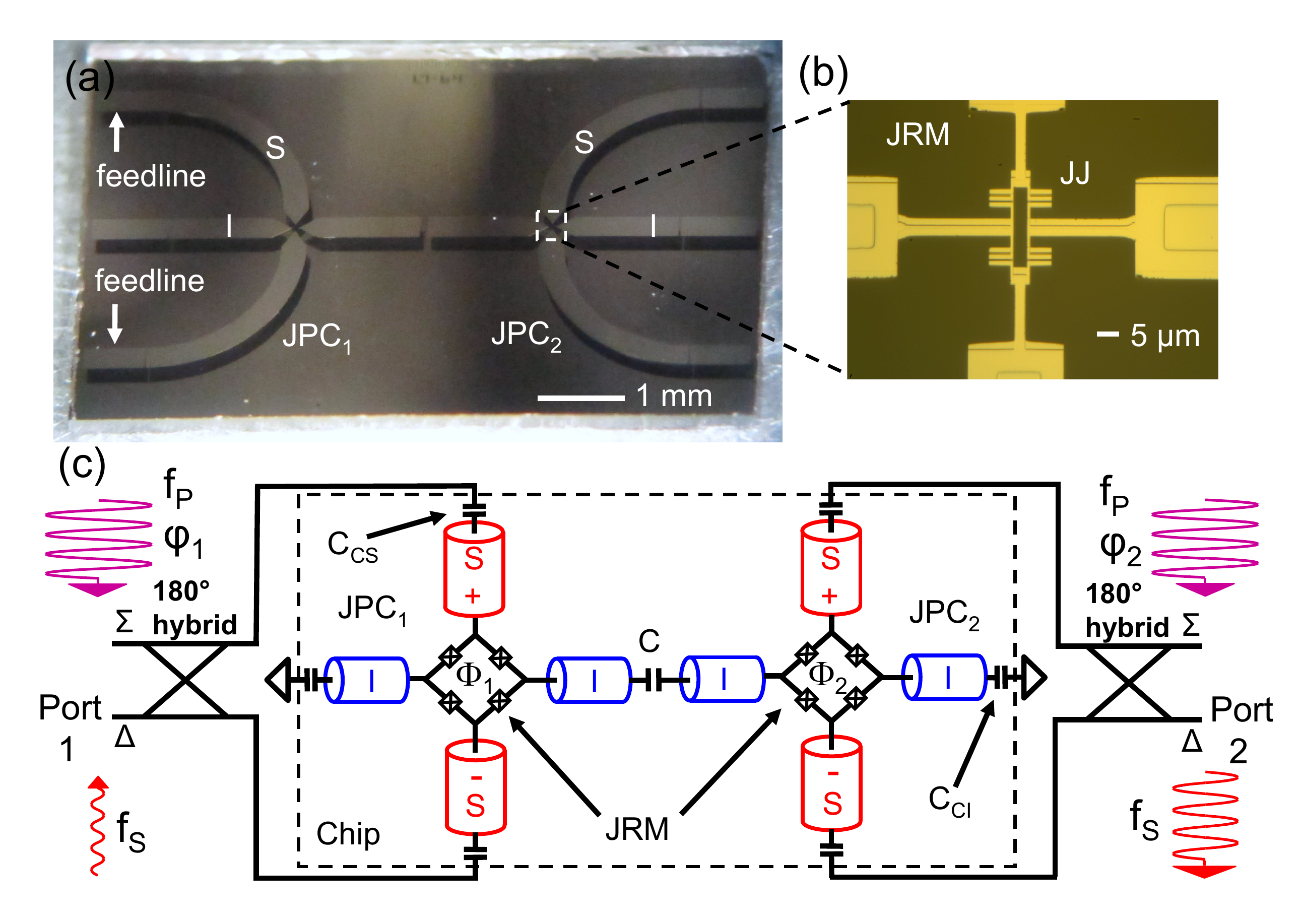}
\caption{(a) An optical micrograph of the device, showing two JPCs
implemented back-to-back on the same chip and coupled through their idler
resonator using a gap capacitor. The chip size is $8.1$ $\operatorname{mm}$
$\times$ $5.7$ $\operatorname{mm}$. The resonators are made of Nb over a $430$
$\operatorname{\mu m}$ thick sapphire substrate. (b) An optical micrograph of one of the JRMs
of the device, which consists of four Al-AlO$_{x}$-Al Josephson junctions. Both JRMs are flux-biased near half a
flux-quantum using two external magnetic coils attached to the copper box
housing the device. (c) A circuit representation of
the two-port device. The input and output signals are fed and measured through the difference ($\Delta$) ports of $180$ degree hybrids connected to  JPC$_{1}$ and JPC$_{2}$. The JPCs are fed with two independent pump tones
at a non-resonant frequency $f_{P}$. The two pumps are injected
through the sum ($\Sigma$) ports of the hybrids and excite a common mode of the
JRM.}%
\label{setup}
\end{center}
\end{figure}

\begin{figure*}
[t!]
\begin{center}
\includegraphics[
width=1.5\columnwidth 
]
{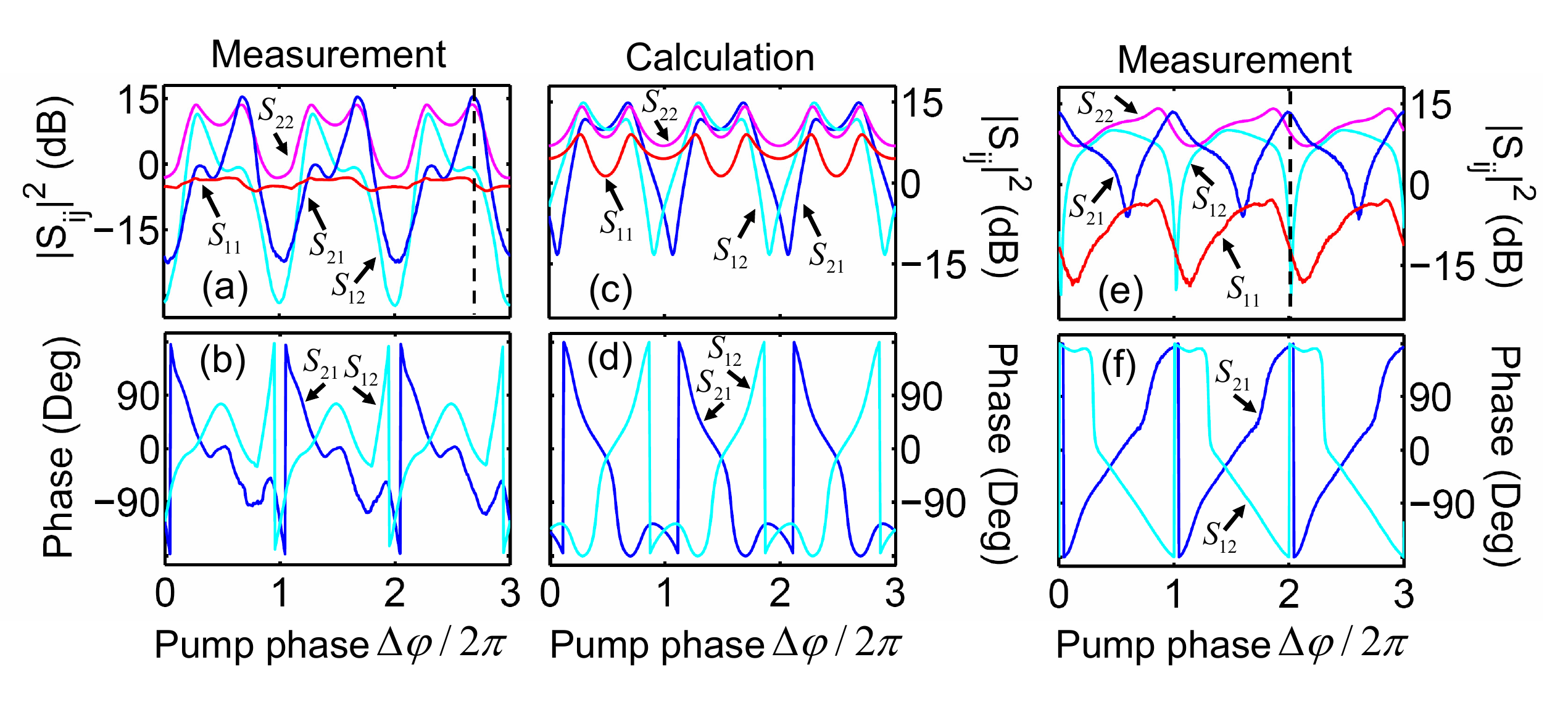}
\caption{Panel (a) exhibits measurements of the scattering parameters of the device, namely $S_{11}$,  $S_{21}$,  $S_{22}$,  $S_{12}$, taken at $f_{S}=8.447\operatorname{GHz}$, as a function of the phase difference between the
pump tones applied at $f_{P}=23.42$ $\operatorname{GHz}$. The vertical dashed black line indicates a desired working point, at which the device functions as a directional amplifier. Panel (b) plots the non-reciprocal phase response of the device measured in transmission $S_{21}$ and $S_{12}$. Panels (c) and (d) exhibit a theoretical calculation of the scattering parameters of the device using the generic signal flow model shown in Fig. \ref{model}. The parameters employed in the calculation are: $\sin\theta_{1}=0.4$, $\sin\theta_{2}=0.8$, $\phi_{1}=\pi/2$, $\phi_{2}=-\pi/2$, $\alpha_{1}=0.6$, $\alpha_{2}=1$, $r_{1}=1.5$, $r_{2}=1.8$. Panels (e) and (f) display an enhanced directional response measured for elevated pump powers. The vertical dashed black line drawn in panel (e) indicates a desired working point. The measurement was taken at $8.468$ $\operatorname{GHz}$ at a different flux bias than the data shown in panels (a), (b). The asymmetric response of $S_{11}$ and $S_{22}$ with respect to the pump phase is attributed to nonlinear processes that arise at relatively high pump drives.
}
\label{meascalc}
\end{center}
\end{figure*}

One example of a superconducting non-reciprocal element is the microstrip SQUID amplifier (MSA)
\cite{MuckSquidHalfGHz,Spietz,MSAnegativeFB,DeFeo,MicrostripDcSquidClarke,Kinion,ArchanadcSQUID}. The MSA is biased with a dc-current and converts an ac flux signal at the
input into an amplified voltage at the output. It suffers from several drawbacks: (1) it dissipates energy
on chip, (2) it has out-of-band back-action which can affect the system under
test, and (3) it is suitable for amplifying low frequency signals of hundreds of
megahertz; engineering a matching circuit so that it amplifies with large enough gain and low enough noise above $6$ $
\operatorname{GHz}$ is difficult.

Another example of a directional element is the traveling wave parametric amplifier \cite{CaltechAmp}. The new amplifier exploits the nonlinear kinetic inductance of superconducting transmission lines in order to parametrically
amplify weak propagating microwave signals. The main disadvantages of this amplification scheme are: (1) it dissipates energy on chip due to the finite resistance of the superconducting films, (2) it does not operate near the quantum limit; the
added noise by the amplifier back to the input is about $3.4$ photons at $9.4$
$\operatorname{GHz}$, (3) it requires an elaborate microwave engineering in order to inhibit
generation of higher harmonics of the pump, limit the amount of dispersion
exhibited by the nonlinearity of the device, and suppress gain
ripples/reflections which arise from imperfect impedance matching, and (4) it lacks temporal and spatial separation between the signal and the pump tones. A new traveling wave propagating parametric amplifier based on Josephson junctions has also been reported very recently \cite{TWPIrfan}. 

In this work, we have implemented and measured a novel Josephson parametric amplifier based on three-wave mixing, which
satisfies the four requirements of directional amplification outlined above.

\section{Concept and implementation}

The basic building block of the new device is the Josephson parametric
converter (JPC), which is a nondegenerate, dissipationless three-wave mixing
amplifier \cite{JPCreview}. The JPC consists of two half-wave microstrip
resonators denoted as signal (S) and idler (I). The two resonators
characterized by a resonance frequency $f_{S}^{res}$ and $f_{I}^{res}$, and an
external quality factor $Q_{S}$ and $Q_{I}$, intersect at a Josephson ring
modulator (JRM), which is positioned at an rf-current anti-node of the
resonators. The JRM consists of four nominally identical Josephson junctions and flux biased with half a flux quantum. When operated in the amplification mode, the device is fed by an intense coherent non-resonant tone denoted as pump (P) at frequency $f_{P}=f_{1}+f_{2}$, where $f_{1,2}$ are the excitation frequencies of the signal and idler (which lie within the bandwidths of the signal and idler resonators). The JRM, which functions as a nonlinear medium, mixes the three waves S, I and P and results in
amplification of the signal and idler due to stimulated downconversion of pump
photons into signal and idler photons. 
By expressing the outgoing wave amplitudes $a^{\mathrm{out}}$, $b^{\mathrm{out}}$ of the signal and idler,
given in units of square root of photon number per unit frequency, as a
function of the incoming wave amplitudes $a^{\mathrm{in}}$, $b^{\mathrm{in}}$
one gets in the frequency domain \cite{JPCreview}%

\begin{align}
a^{\mathrm{out}}\left[  +\omega_{1}\right]    & =ra^{\mathrm{in}}\left[
+\omega_{1}\right]  -i\sqrt{r^{2}-1}e^{-i\varphi}b^{\mathrm{in}}\left[
-\omega_{2}\right], \label{aout}\\
b^{\mathrm{out}}\left[  -\omega_{2}\right]    & =rb^{\mathrm{in}}\left[
-\omega_{2}\right]  +i\sqrt{r^{2}-1}e^{i\varphi}a^{\mathrm{in}}\left[
+\omega_{1}\right] , \label{bout}
\end{align}
where $r=\sqrt{G}$ is the amplitude reflection parameter, $G$ is the power
gain of the device at the working point, $\varphi$ is the phase of the pump, and $\omega_{1,2}=2\pi$$f_{1,2}$. The first and second terms on the right-hand side of Eqs. (\ref{aout}), (\ref{bout})
correspond to a cis-gain and trans-gain processes respectively. While the cis-gain
process relates between output and input fields at the same port and
frequency, the trans-gain process relates fields separated by frequency $f_{P}$ across different ports. Note that the phase acquired in the trans-gain process is non-reciprocal and depends on the phase of the pump.

The new device shown in Fig. \ref{setup} consists of two nominally
identical JPCs paired together back-to-back on the same chip. The JPC$_{1}$ on the left and JPC$_{2}$ on the right, are coupled via a gap capacitor $C\simeq20$ f$
\operatorname{F}
$ between the two idler resonators, whose resonance frequency is $f_{I}^{res}=15 \pm 0.1$ $
\operatorname{GHz}
$. The other end of the idler resonators is capacitively coupled to a
superconducting island which is wire-bounded to ground. Thus, the idler
eigenmode is inaccessible through external ports and
functions as an internal mode of the combined system. The device has
two ports (see Fig. \ref{setup} (c)), the feedlines of the S resonators of JPC$_{1}$ and JPC$_{2}$. At the flux bias working point, the S resonators of JPC$_{1}$ and JPC$_{2}$ resonate at $f_{S1}^{res}=8.47$ $
\operatorname{GHz}
$, $f_{S2}^{res}=8.45$ $
\operatorname{GHz}
$ with a bandwidth of $\kappa_{S1}/2\pi=122$ $
\operatorname{MHz}
$, $\kappa_{S2}/2\pi=116$ $
\operatorname{MHz}
$ respectively, which correspond to a total quality factor of $Q_{S1}=69$, $Q_{S2}=73$.
Similar to the excitation scheme of a single JPC, the S resonators are addressed through the difference ($\Delta$) port of a $180$ degree hybrid. The two JPCs are driven by two independent non-resonant pump tones. The pumps share the same frequency $f_{P}$, but can have different phases $\varphi_{1}$,
$\varphi_{2}$.

The main idea of the device is based on utilizing the non-reciprocal phase response of the
JPC in the amplification mode in conjunction with wave interference between multiple paths. These paths are formed, in the device, as a result of the couplings that exist
between the idler $I_{1}\leftrightarrow I_{2}$ and signal $S_{1}\leftrightarrow S_{2}$ resonators of the two JPCs. The latter coupling is mediated by a ground plane mode, that is enhanced by the idler conductor between the two resonators. Such direct coupling has been experimentally observed by injecting a weak signal
though port $1$ and measuring the transmission through port $2$ (and vice
versa) without applying any pump tones. It has also been verified using microwave
simulations of the device configuration. As a result of the two couplings
$I_{1}\leftrightarrow I_{2}$ and $S_{1}\leftrightarrow S_{2}$, feedback loops of the amplified signals
are formed between the two JPC stages. Of special importance in this scheme, is the trans-gain process in which signal photons are upconverted to idler and downconverted to signal with gain ($S_{1}\leftrightarrow I\leftrightarrow
S_{2}$). Due to the non-reciprocal phase response of the JPC, a signal that
undergoes sequential trans-gain processes acquires a
non-reciprocal phase shift which depends on the relative phase difference
between the pumps. In one direction, it would acquire a phase shift of
$\left(  \varphi_{1}-\varphi_{2}\right)  $, whereas in
the other direction, it would acquire an opposite phase $-\left(  \varphi
_{1}-\varphi_{2}\right)  $. In contrast, amplified signals which follow a path
of direct coupling acquire, regardless of direction, a constant phase shift
which depends solely on the coupling parameters.

\section{Directional amplification}
    In Fig. \ref{meascalc} (a) and (e) we plot power gain measurements for the four scattering parameters of the device, i.e. $S_{11}$, $S_{21}$, $S_{12}$, $S_{22}$, as a function of the relative phase difference between the applied pumps at $f_{P}=23.42$ $\operatorname{GHz}$. The gain plots, obtained for different flux biases, are measured at $f_{S}=8.447$ $ \operatorname{GHz}$ and $8.468$ $ \operatorname{GHz}$ respectively. In the latter case of panel (e), the resonance frequencies of the two JPCs are $f_{S1}^{res}=8.44$ $\operatorname{GHz}$ and $f_{S2}^{res}=8.43$ $\operatorname{GHz}$. In both measurements, the two pump generators are phase-locked to the $10$ $\operatorname{MHz}$ reference oscillator of a rubidium atomic clock. The phase difference between the pumps is varied in a continuous manner as a function of time, by introducing a frequency offset of a few hertz between the two generators. As can be seen in panel (a) (panel (e)), for a certain relative pump phase difference, indicated by a vertical dashed line, the device amplifies input signals in transmission $S_{21}$ with a maximum power gain of $15$ dB ($13.6$ dB) and attenuates reflections $S_{11}$ by $4$ dB ($12.5$ dB). On the other hand, incoming signals in the opposite direction are attenuated in transmission $S_{12}$ by $2$ dB ($19$ dB) and amplified in reflection $S_{22}$ by $13$ dB ($9.5$ dB). Note that the amplification in reflection ($S_{22}$), despite being generally undesirable, does not influence port 1 (which would potentially be connected to a quantum system). The main differences between the two measurements are, (1) the data in panel (e) displays, at the desired working point, a significantly stronger directional response than panel (a), i.e. about $32$ dB difference between $S_{21}$ and $S_{12}$ in panel (e) versus $17$ dB in panel (a), and (2) the reflection curves $S_{11}$ and $S_{22}$ in panel (e) exhibit an asymmetrical response with respect to the relative pump phase compared to the curves in panel (a). We attribute the observed asymmetry to nonlinear processes that arise at relatively high pump drives (about 3 dB higher on both pumps), which are applied in the case of panel (e) compared to panel (a). 
    
\section{ Theoretical Model}

\begin{figure}
[htb]
\begin{center}
\includegraphics[
width=\columnwidth 
]
{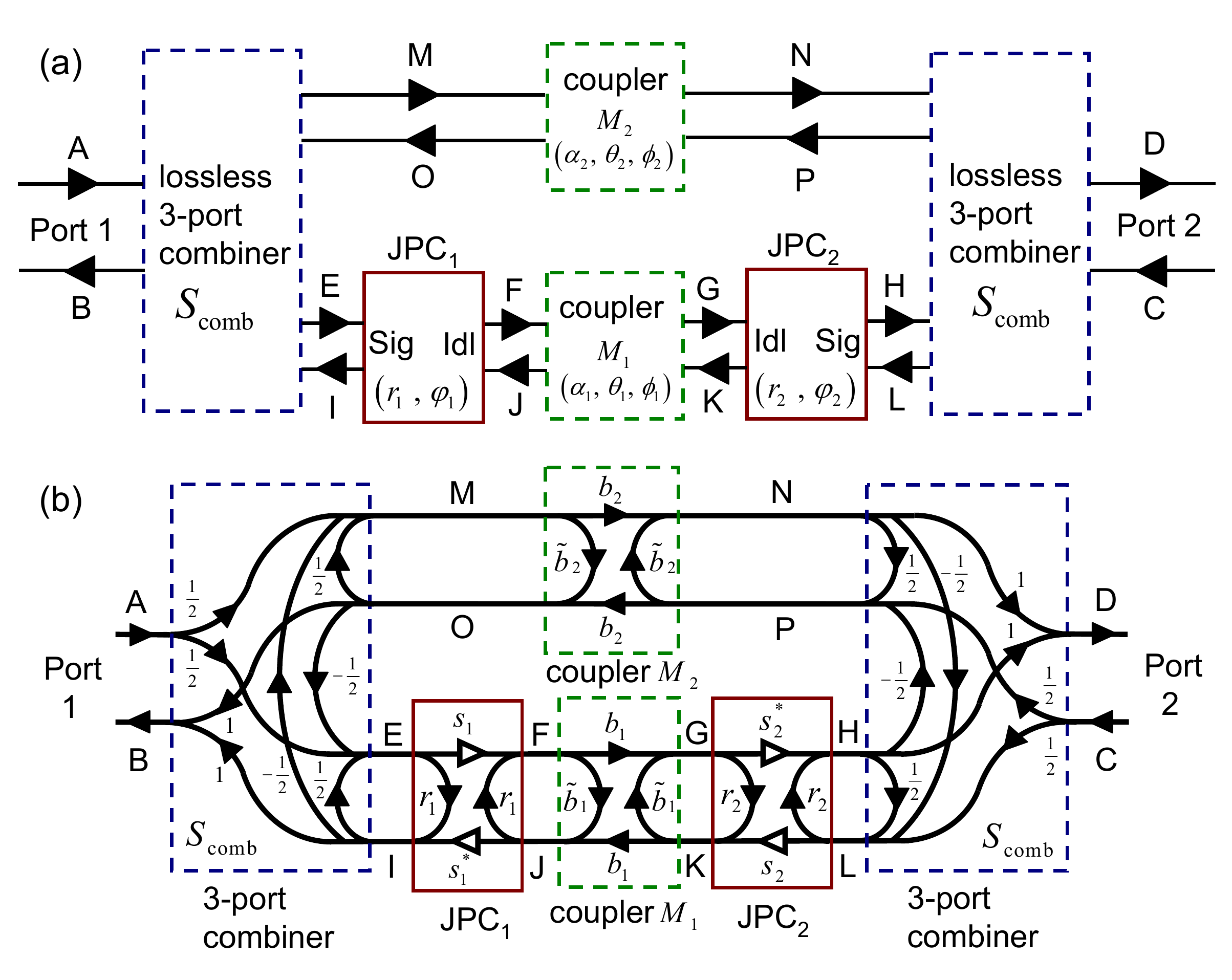}
\caption{(a) A generic signal flow graph modeling the system. The block diagram consists of two JPCs (represented by solid red rectangles) coupled through their idler and signal ports by effective 2-port couplers (beam-splitters) $M_{1,2}$ (represented by dashed green rectangles). The additional coupling between the signal ports of the device and its external ports 1 and 2 is established by introducing effective 3-port combiners/splitters (represented by dashed blue rectangles). (b) A detailed version of the signal graph depicted in panel (a). This graph is used in the derivation of Eqs. (\ref{E})-(\ref{P}) and (\ref{D})-(\ref{B}) of the theoretical model. It is also used in the calculation of the device response shown in panels (c) and (d) of Fig. \ref{meascalc}.}
\label{model}
\end{center}
\end{figure}

In panels (c), (d) of Fig. \ref{meascalc}, we plot a theoretical calculation based on a generic model of the
device schematically drawn in Fig. \ref{model} (a). In this model, the two JPCs are effectively coupled together through 2-port couplers (beam-splitters) $M_{1,2}$, where $M_{1,2}$ connect between the idler and signal ports respectively. The scattering matrix of the 2-port couplers $M_{1,2}$ is of the form 

\begin{equation}
M_{1,2}=\alpha_{1,2}e^{i\phi_{1,2}}\left(
\begin{array}
[c]{cc}%
\cos\theta_{1,2} & i\sin\theta_{1,2}\\
i\sin\theta_{1,2} & \cos\theta_{1,2}%
\end{array}
\right)  , \label{M12}
\end{equation}

where the angles $\theta_{1,2}$ set the reflection and transmission amplitudes
of each coupler, $\phi_{1,2}$ are global phases, and $0<\alpha_{1,2}\leq1$ are
loss factors (for $\alpha_{1,2}=1$ the couplers are lossless). In addition, in
order to account for the coupling between the external ports 1 and 2 of the device and the signal
port of each JPC, we assume for simplicity, a lossless
3-port combiner/splitter whose scattering matrix is given by

\begin{equation}
S_{\mathrm{comb}}=\left( 
\begin{array}
[c]{ccc}%
0 & 1 & 1\\
1/2 & 1/2 & -1/2\\
1/2 & -1/2 & 1/2
\end{array}
\right)  , \label{Scomb}
\end{equation}
where the scattering parameters of the first row determine the generated output field on
the external port 1 or 2 of the device.

To obtain the theoretical curves, amplitude and phase, plotted in Fig. \ref{meascalc} (c), (d), we write the signal flow equations corresponding to the nodes E to P, indicated in Fig. \ref{model} (b), using Eqs. (\ref{aout})-(\ref{Scomb}),

\begin{align}
E-\frac{1}{2}I+\frac{1}{2}O  & =\frac{1}{2}A,\label{E}\\
F-s_{1}E-r_{1}J  & =0,\label{F}\\
G-b_{1}F-\tilde{b}_{1}K  & =0,\label{G}\\
H-s_{2}^{\ast}G-r_{2}L  & =0,\label{H}\\
I-r_{1}E-s_{1}^{\ast}J  & =0,\label{I}\\
J-b_{1}K-\tilde{b}_{1}F  & =0,\label{J}\\
K-s_{2}L-r_{2}G  & =0,\label{K}\\
L-\frac{1}{2}H+\frac{1}{2}N  & =\frac{1}{2}C,\label{L}\\
M-\frac{1}{2}O+\frac{1}{2}I  & =\frac{1}{2}A,\label{M}\\
N-b_{2}M-\tilde{b}_{2}P  & =0,\label{N}\\
O-\tilde{b}_{2}M-b_{2}P  & =0,\label{O}\\
P+\frac{1}{2}H-\frac{1}{2}N  & =\frac{1}{2}C,\label{P}%
\end{align}
where 
\begin{equation}
s_{1,2}=-i\sqrt{r_{1,2}^{2}-1}e^{-i\varphi_{1,2}},\label{s}%
\end{equation}
and
\begin{align}
b_{1,2}  & =i\alpha_{1,2}\sin\theta_{1,2}e^{i\phi_{1,2}},\label{b}%
\\
\tilde{b}_{1,2}  & =\alpha_{1,2}\cos\theta_{1,2}e^{i\phi_{1,2}},\label{btelde}
\end{align}
we also express the output nodes $D$ and $B$ as %
\begin{align}
D  & =H+N,\label{D}\\
B  & =O+I.\label{B}
\end{align}

By solving the amplitude equations for nodes $E..P$ in response to a unity
input signal $A=1$ and $C=0$, we obtain the scattering parameters $S_{21}$ and
$S_{11}$ for outputs $D$ and $B$ respectively. Similarly, by solving the same
equations for a unity input signal $A=0$ and $C=1$, we obtain the scattering
parameters $S_{22}$ and $S_{12}$ for outputs $D$ and $B$ respectively. 

Despite the simplicity of the theoretical model which we use to describe the complex dynamics of the system, the calculated curves show a good qualitative agreement with most of the scattering parameters data shown in Fig. \ref{meascalc} (a) and (b). The main features which are successfully captured by the theoretical model are, (1) the non-reciprocity of the amplitude and phase response of the transmitted signals, $S_{21}$ versus $S_{12}$, (2) the order of magnitude of the scattering parameter amplitudes, (3) the symmetrical response of the scattering parameters with respect to the pump phase difference, and (4) the fact that the reflection parameters $S_{11}$ and $S_{22}$ can be different. On the other hand, the model falls short in yielding exact quantitative agreement with the data shown in panel (a), and in reproducing the asymmetric curve shapes of the scattering parameters, especially $S_{11}$ and $S_{22}$, shown in panel (e) (even for different values of the model parameters). While the former shortcoming can be partly explained by the large parameter space of the theoretical model, which we tried to constrain as much as possible, we attribute the latter shortcoming to nonlinear effects, which come into play at elevated pump powers and are not accounted for in this model. 

 In addition to explaining the data, it is worthwhile pointing out that solving the theoretical model gives some insight to certain requirements for directional amplification. In particular, we find that (1) a finite amount of ``loss" between the two JPC stages is required in order to obtain gain and directionality. A possible candidate for loss in our system is power leakage to other modes. For example, leakage to higher order modes of the system, or to the even mode, due to finite imbalance between the even and odd modes. (2) Some imbalance between the reflection parameters of the two stages $r_{1}$ and $r_{2}$ is needed in order to have different $S_{11}$ and $S_{22}$ responses. This finding agrees with the experiment since the observed gains in reflection on the two stages, when pump tones are applied separately, are generally unequal (about 9 dB on one stage and 2 dB on the other). (3) Only certain coupling strengths and acquired phases between the two JPCs, give rise to directional amplification. This result is in agreement with the experiment as well, since we are only able to measure directional amplification for certain applied fluxes in the JRMs and pump frequencies. The former control ``knob'', the magnetic flux, sets the offset between the resonance frequencies of the two stages, while the latter, the pump frequency, determines the range of frequencies $f_{1}$ and $f_{2}$ which get excited within the bandwidths of the S and I resonators.  
   
\section{Device Performances}

Having shown in Fig. \ref{meascalc} that the device satisfies requirements (1), (2), and (3) of directional amplification, namely, the requirements of transmission, input-matching and non-reciprocity, we now set bounds on the device performances, i.e. dynamical bandwidth, added noise and maximum input power.

\begin{figure}
[htb]
\begin{center}
\includegraphics[
width=\columnwidth 
]
{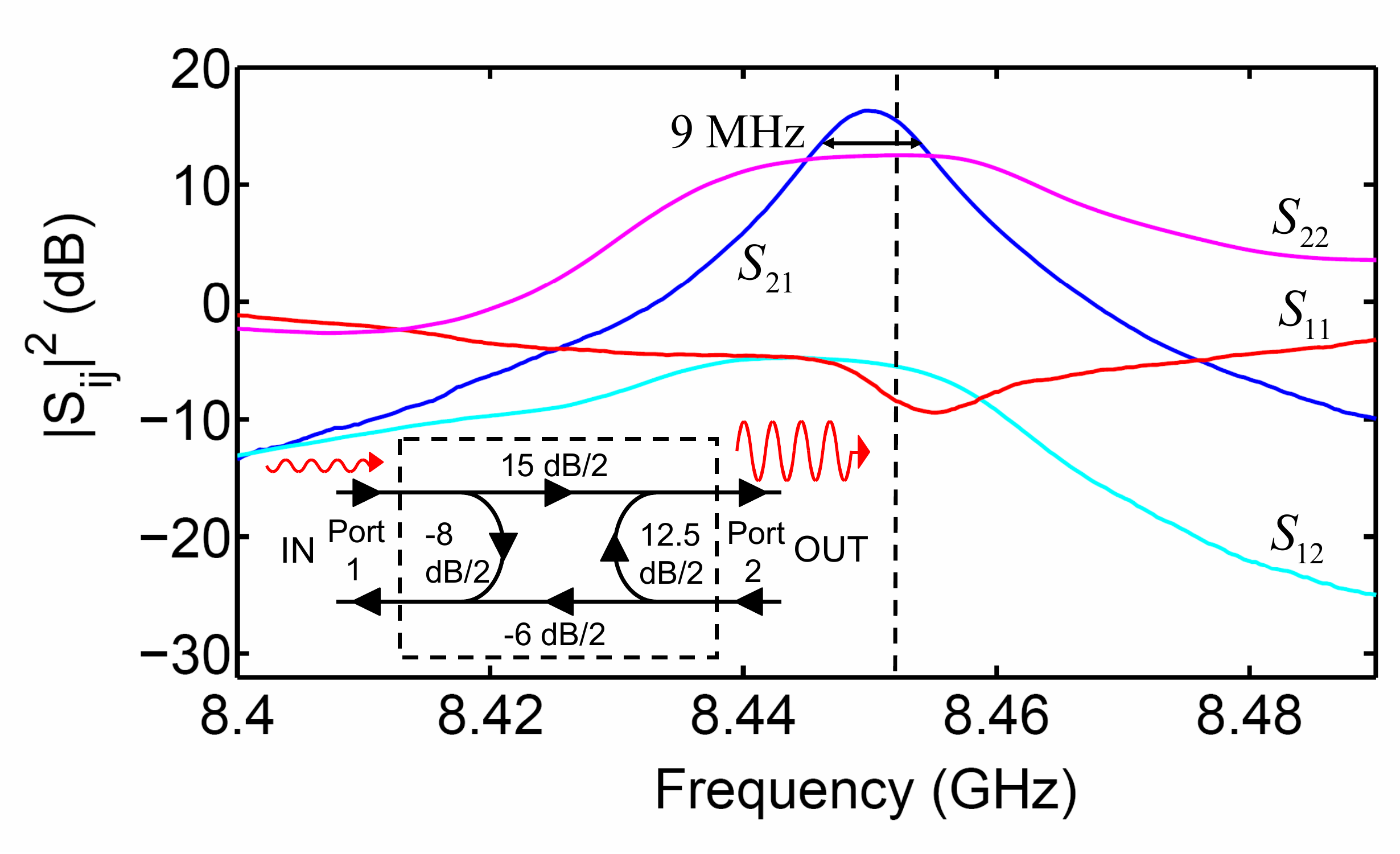}
\caption{A measurement of the scattering parameters of the device as a function of signal frequency taken for a fixed pump phase difference which maximizes directionality. The dynamical bandwidth of the device measured at the 3 dB points below the maximum $S_{21}$ gain is 9 $ \operatorname{MHz}$. In the bottom left part of the figure we display a schematic sketch of the amplitude gain of the device at $f_{S}=8.453$ $\operatorname{GHz}$, which corresponds to the dashed vertical line. The pump amplitudes applied in this measurement are the same as in Fig. \ref{meascalc} (a).}%
\label{gainfreq}
\end{center}
\end{figure}

In Fig. \ref{gainfreq}, we plot the device scattering parameters measured as a function of signal frequency at the desired working point, indicated by a vertical dashed line in Fig. \ref{meascalc} (a). We find that the dynamical bandwidth of the device $9$ $
\operatorname{MHz}$ taken at a power gain of $16$ dB, is mainly limited by the bandwidth of the signal resonators. Such dynamical bandwidth is, in general, suitable for qubit-state readout applications, as it exceeds the bandwidths of most readout cavities. Nevertheless, for any practical qubit application, some tunability is required in order to easily match between the center frequency of the amplifier and the readout frequency. One possible way to achieve that, is by substituting the present JRM scheme (shown in Fig. 1 (b)) by an inductively shunted version similar to the one introduced in Ref. \cite{Roch}. Incorporating such rings in next generations of our device is expected to extend the tunable bandwidth to more than $100$ $\operatorname{MHz}$.

Another important figure of merit for quantum signal amplification is added noise. By measuring the improvement in the signal-to-noise ratio of the system due to the amplifier and using our knowledge of the noise temperature
of the output chain, we are able to set an upper bound on the noise added to
the input by the amplifier. Using this measurement method, we find that the
device adds up to $1\pm0.5$ photons at the signal frequency which is similar
to the bound we get for single JPC amplifiers.

\begin{figure}
[htb]
\begin{center}
\includegraphics[
width=\columnwidth 
]
{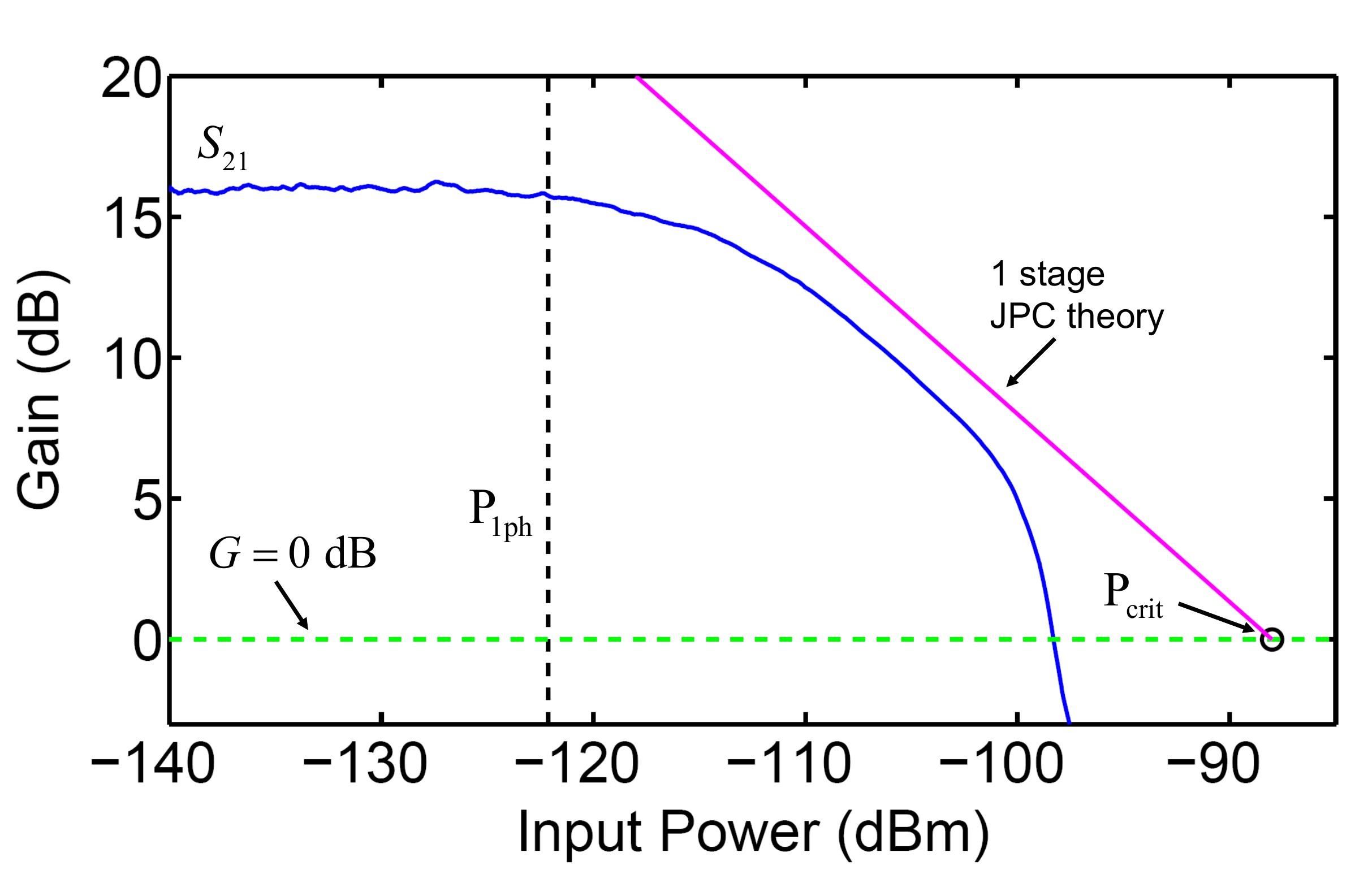}
\caption{A measurement of the maximum input power of the device obtained at the desired working point, indicated by a vertical dashed line in Fig. \ref{meascalc} (a). The power gain $S_{21}$ drawn in blue, is measured as a function of the input signal power. The vertical dashed black line indicates the input power of one photon at the signal frequency per inverse dynamical bandwidth at $G=16$ dB. The solid magenta line is the expected upper limit on the maximum input power of a one-stage JPC due to pump depletion effect \cite{JPCreview}.}
\label{dr}
\end{center}
\end{figure}

Furthermore, in Fig. \ref{dr} we display a maximum input power measurement taken for $S_{21}$ at the desired working point indicated in Fig. \ref{meascalc} (a). The device gain, drawn in blue, is measured as a function of the input signal power. This measurement shows that the device at its maximum gain of $16$ dB can tolerate at least three input photons at the signal frequency per inverse dynamical bandwidth. It also shows that the limiting mechanism on the maximum input power is pump depletion. This can be seen by comparing the gain curve to the magenta line, which depicts the bound on the input power of one stage JPC due to this effect \cite{JPCreview}. A possible improvement of this figure of merit is discussed in Refs. \cite{Jamp,JPCreview}.

\section{Summary}
We have implemented and measured a proof-of-principle, directional
Josephson amplifier suitable for qubit readout, which employs a novel
non-reciprocal mechanism on chip and does not involve magnetic materials. To
enhance the device performances, a deeper understanding of the role played by the microwave couplings between the two stages is required. An improved version of this amplifier with respect to microwave control and tunable
bandwidth could allow in the future on-chip integration with qubit-cavity systems.

\begin{acknowledgements}
Discussions with R. J. Schoelkopf, A. Kamal, A. Roy, F.
Schackert, and M. Hatridge are gratefully acknowledged. The assistance of
M. Power in the fabrication process is highly appreciated. This research
was supported by the NSF under grants DMR-1006060 and DMR-0653377; the NSA through ARO Grant
No. W911NF-09-1-0514, IARPA under ARO Contract No. W911NF-09-1-0369. M.H.D. acknowledges partial support from College de France.
\end{acknowledgements}

\appendix
\section*{Appendix A: Fabrication and Measurement Details}

The device resonators are made of Nb over a $430$
$\operatorname{\mu m}$ thick sapphire substrate. A $2$ $\operatorname{\mu
m}$ thick silver ground plane is e-beam evaporated on the back side of the substrate to enhance
thermalization and microwave control. The Nb layer was patterned using a
standard photolithography step and etched using reactive ion etching. The two JRMs
of the device were incorporated using a standard e-beam lithography process
followed by two angle shadow evaporation of aluminum (with an oxidation step
in between) and lift off. A large overlap area (partially shown in Fig. 1) is established between the Nb part of the resonators
and the Al wires of the JRM which was preceded by plasma cleaning. No measurable losses were observed in our samples due to this interface. The Josephson junction area is $5\operatorname{\mu m}\times1\operatorname{\mu m}$,
while the loop area of the JRM is about $50\operatorname{\mu m}^{2}$. The
critical current of the nominally identical Josephson junctions of the JRM is
$I_{0}=3\pm0.5\operatorname{\mu A}$.

The measurements are taken in a dilution fridge at a base temperature of $30
$ m$
\operatorname{K}
$. The experimental setup used is similar to those described in Refs. \cite{Jamp,BSconv} with a few differences. For example, in the present setup, we use two independent pump lines, and two magnetic coils. We also do not have an idler line (input and output), instead we use two signal lines (input and output) connected to port 1 and port 2 of the device.  

The measurement shown in Fig. 2 is taken using a two-port vector network analyzer (VNA) operated in a zero frequency span mode centered at $f_{S}$. In order to measure the dependence of the scattering parameters on the relative phase between the two pump tones, we introduce a frequency offset of $10\operatorname{Hz}$ between the pump generators and measure $1\operatorname{s}$ duration time sweeps on the VNA. The measured time sweeps are externally triggered using an arbitrary wave generator with a periodic signal of $1\operatorname{Hz}$ and yield, as expected, $10$ periods of the scattering parameter as a function the pump phase difference (in Fig. 2 we show only three periods). To ensure a rapid measurement of the different scattering parameters of the device (in order to avoid possible phase drifts between the two microwave generators over time) a room-temperature 2-way switch is installed on each port of the VNA. Thus, allowing fast switching between the input of JPC$_{1}$ and JPC$_{2}$, connected to port 1 of the VNA, and between the output of JPC$_{1}$ and JPC$_{2}$, connected to port 2. In all measurements, the generators and measuring devices are phase-locked to the same $10\operatorname{MHz}$ reference of a rubidium atomic clock.

\end{document}